# Correlation of Scholarly Networks and Social Networks

Amir Razmjou

Florida Institute of Technology (FIT), United States

`arazmjou2014@my.fit.edu`

**Abstract.** In previous studies, much attention from multidisciplinary fields has been devoted to understand the mechanism of underlying scholarly networks including bibliographic networks, citation networks and co-citation networks. Particularly focusing on networks constructed by means of either authors' affinities or the mutual content. Missing a valuable dimension of network, which is an audience scholarly paper. We aim at this paper to assess the impact that social networks and media can have on scholarly papers. We also examine the process of information flow in such networks. We also mention some observations of attractive incidents that our proposed network model revealed.

## 1 Introduction

It is evident that science and technology have always had a prominent impact on society particularly at the age of communications that impact is growing exponentially. But does the characteristics of our society and substantial role that media and social networks also effect on the direction of science and technology? Can we take the fact for granted that science is the only force responsible for achievement and discoveries in science? Is that still plausible to examine the importance of scholarly works in isolation of contribution key player, reader, of that work? In this study we proposed a novel scholarly network to proliferate role of social networks and researchers attentions in scholarly interactions.

Social Networks along with new developments in "Network Science" enable researcher to gain more visibility about human interactions. On the other hand scholarly network analysis presents scientists and policy makers set robust tools to support their decisions [1]. In this study we propose a novel network definition that incorporates audiences of scholarly works. This novel approach opens up a new perspective that is, social network users can be potential factor for information flow and thus demotion scholarly networks structure. This approach is in contrast to coauthorship, co-citation, bibliographic coupling, or co-words networks. This study further explores examines proposed network structures both in terms of micro-level to community structures. We also correlate this network with citation network and validate our findings with our experimental data.

Social networks often can be modeled into Complex Networks that have topological and properties of scale-free networks with huge number of nodes and edges. Analysis of these of networks a daunting task without having right toolset at hand. One promising approach is to decompose the networks into smaller sub-networks, namely network communities. Classification of communities in network assists to uncover better understanding of each component of networks individually. In addition, the resulting network of communities can give us a border perspective of overall network structure.

## 2   Related Works

Recently a lot of literature devoted on impact of social networks on diverse aspects of society. Considering that society, economy and social networks can influence scholarly networks. Gunther Eysenbach claims "Tweets can predict highly cited articles within the first 3 days of article publication". And also discusses impact of social media on citations improvements but also points out that the true use of these metrics is to measure the distinct concept of social impact and he proposed social impact measures based on tweets only as complement conventional citation metrics" [1]. Other studies regarding social networks show that long-term analysis of the popularity topic can present two peaks in the buzzwords. The first peak is significantly inferior to the second one but these two peaks can represent different diffusions of the same word, in which the former only reaches a small scope while the latter not only spreads globally but also explains for long-lasting diffusion across the network [2]. Another study takes multi-dimensional approach to compare different types scholarly networks similarity.They find topical networks[1] and coauthorship networks have the lowest similarity and these two sets of networks categorize two sets of networks with the high similarity, co-citation networks and citation networks on one hand and co-word networks and topical networks on the other hand [3].

---

[1] Topical network refers to "the collection of sites commenting on a particular event or issue, and the links between them" (Highfield, Kirchhoff, & Nicolai, 2011, p. 341).

We defined Social Scholarly Network (SSN) terms of a social network can support connection two papers in that perspective. In SSN nodes represent papers as conventional scholarly network and edge between two papers exists if two papers are mentioned by the same user and weight for that edge is proportional to the time interval that user mentioned these papers.

Three basic metrics are constructed. One is user participation in promoting paper in social network $U_{ij} = 1$ if paper $j$ is mentioned by user $i$. Second $C_{ij} = 1$ if paper $j$ is co-authored with paper $i$. In our undirected SSN network we calculated the weight of edges

$$W_{ij} = \begin{cases} \frac{1}{\sqrt{T_{ij}}} \cdot RT, & T_{ij} < TimeWindow \\ 0, & T_{ij} \geq TimeWindow \end{cases}$$

In which $W_{ij}$ weight edge connecting $i$ node to $j$ is disproportional to time interval that user tweeted $i$ and $j$ papers and proportional to $RT$, which is the normalized *Retweets* count for that paper.

A multi-relational network is composed of two or more sets of edges between a set of vertices. A multi-relational network can be defined as $M = (V, E)$, where $V$ is the set of vertices in the network, $\mathbb{E} = \{E_1, E_2, \ldots E_m\}$ is a family of edge sets in the network, and any $E_k \subseteq (V \times V): 1 \leq k \leq m$.

For each edge set in $\mathbb{E}$ there's, or categorical or dimensional. For example, within the same network $M, E_1 \text{ and } E_2 \in \mathbb{E}$ may denote "Tweeted by the same user" and "coauthor ship," respectively.

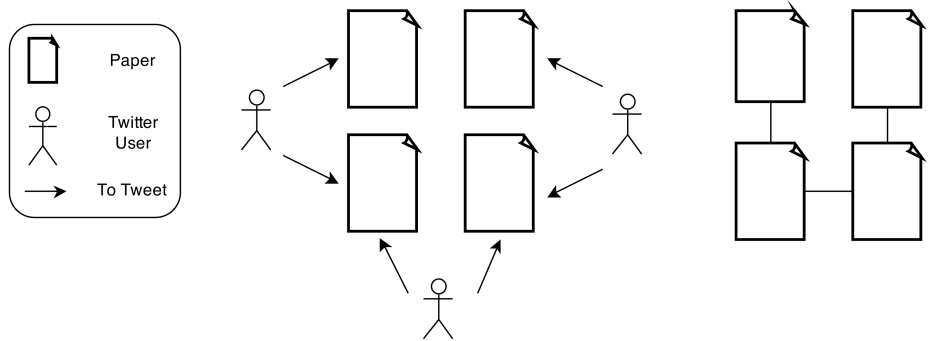

**Figure 1.** Structure of social citation network incorporating tweeter user roles to infer papers relations.

## 3  Data Collection

Before describing the dataset we used in this evaluation it is worth going into some detail on how Tweeter tracks embedded shortened Links. The goal for tweeter is to have both versions of shortened link as 140 characters policy. On the other hand possession of lengthened link on search database to increase their user search experience so users will be able to search the links no matter of shortening process. But there's a limitation on how far tweeter will risk resources to lengthen a link by Firstly, there's a limitation in order to prevent bots to not fall into cyclical links. Second long links won't be stored in tweeter database as it might make system vulnerable in term of storage resources.

In this study we have gathered data from Twitter using REST API. Simply by searching for *arXiv.org* keyword. Using cloud services we managed to collect ~16000 tweets. This number didn't account for RTs but we took number of RT in consideration when specifying weights for our network edges. After making sure that every link is already lengthened and points to a specific page on arXiv.org web site. We were able to add furthermore *arXiv.org* metadata, these additional attributes enriched our notion of papers mentioned in *arXiv.org* by means of Published Date, Authors, Title, Summary, Category by calling *arXiv.org* REST API and adding them to our `MongoDB` documents we have collected so far. In order to complement our dataset with "Citation Number" we crawled Google Scholar for each paper. In addition of citation number we obtained Google Scholar Cluster ID that is, the cluster identification to find similar papers. Once we constructed our network we came up with huge star nodes these nodes happened to be tweeter controlled bots. We also managed to eliminate bot users by counting degree distribution and eliminating nodes with extraordinary high degree. Using authors' identifications we obtained from *arXiv.org* we constructed an entirely different set edges by means of citation network by iterating over authors and co-authors and mining relationships. We also eliminated bot controlled twitter users by finding the deviation in density of posts in a time span.

We applied time-window restriction over all relating papers to eliminate formation of network cliques but a fixed-size time-window animalizes distribution of edges in different communities by skewness it applies to our data. That's due the fact that users with different social behaviors and habits tend to have a relatively diverse rate of social network interactions.

By applying network modularity algorithm using different resolution values we formed communities. Then we compared later relation-aware method with relation-agnostic methods of clustering on node attributes. We used different set of attributes like paper category, subcategory and compound value of both.

## 4    Results

*Scale-free networks* are a type of network characterized by the presence of large hubs. A scale-free network is one with a power-law degree distribution. For an undirected network, we can just write the degree distribution as

$$P_{deg}(k) \propto k^{-\gamma} \quad (2)$$

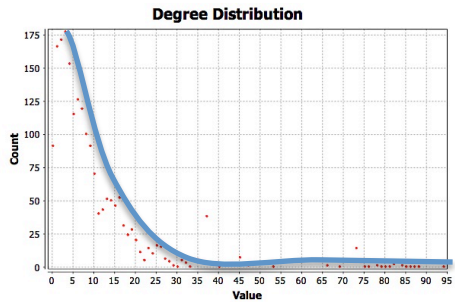

**Figure 2**. Degree Distribution obtained from social scholarly network matches power-law degree distribution.

We begin by analyzing small part of our network by refining our nodes restricted to papers in regard to Computer Science.

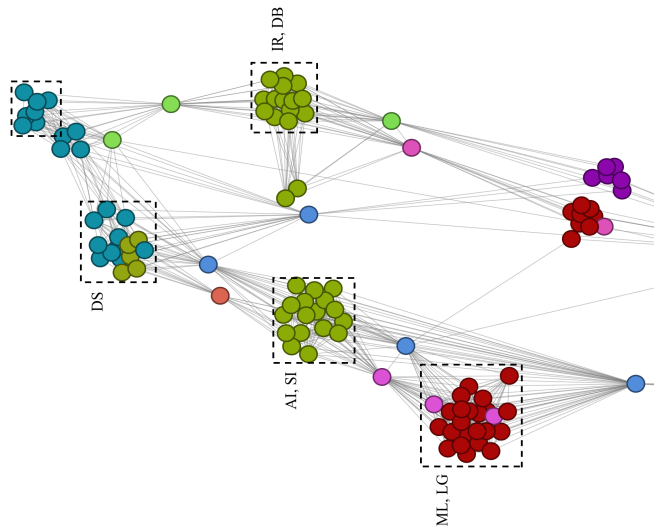

**Figure 2.** Diversity of sub-category in a single community. As expected two related subcategories like Information Retrieval and Databases emerged in a same community (IR, DB). As our modularity algorithm is insulated from these attributes of nodes this evidence validates the correlation of nodes in our social scholarly network. By reviewing user profiles of AI, SI community we discovered that most users in this community remarkably are interested in social networks data mining. The same reasoning is not true about DS, data structure community. Another interesting outcome of this network is that the single nodes between different communities are seem to be interdisciplinary paper citing all materials in different communities they are connected to

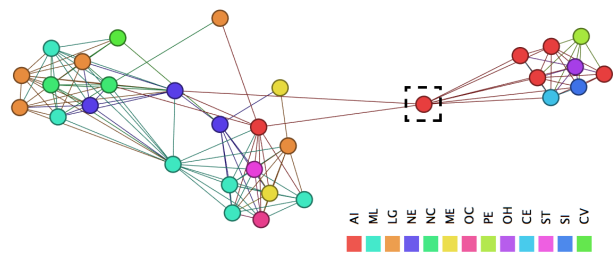

**Figure 3.** Demonstrates ensemble of different fields from Computer Science. The marked paper "The Limitations of Standardized Science Tests as Benchmarks for Artificial Intelligence Research: Position Paper" discusses about regents tests a subject that is applicable. The citation number of 1512 for this paper is almost one of highest citation number among Computer Science nodes in our network, betweenness of ~8.2 and RT number of larger than 100 indicate that multi-disciplinary papers is only well-standing among citation network but shows a great appreciation from social network users. Another interesting founding is that degree numbers of all bridge nodes, in terms of the number of communities they connect together, are identical and equal to two. That also implies that even multi-disciplinary papers cite no more than two diverse fields at the same time.

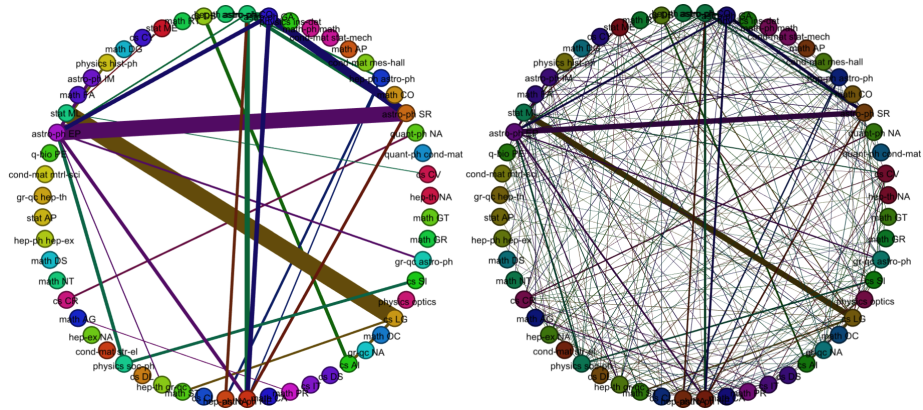

**Figure 4.** Small world phenomena of network of communities. Top network represents connection of communities in a week long time-window, Bottom shows most dominant edges.

By grouping different type of categories of science in network and aggregating the connections. We obtained a network with small world phenomena. Our network exhibits diameter of 5 and average length path of 2.28. The small world phenomena once again validated tendency of social network users to combine diverse disciplines together. By concentrating on more dominant connections it appears that "Statistical Machine Learning" (stat-ml) and computer science machine learning is extremely connected which is expected. But relatively dominance of "Cryptography and Security" and "Quantum Physic " is of interest. It seems at 12 March 2014, Stephanie and Nelly Ng researchers at the Institute for Quantum Computing (IQC) at the University of Waterloo, Canada managed to perform random oblivious transfer of 1,366 bits in less than three minutes and that resulted in trend in social networks at the time we gathered our data. Another interesting aspect is relevance of can be found by affiliation of "Physics and Society" to "Computer Science Social and Information" (physics.soc-ph and cs.si).

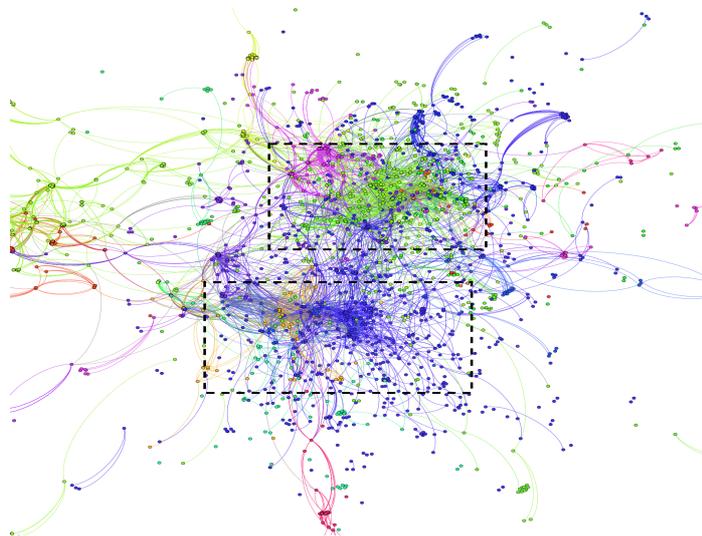

**Figure 5.** Concentration of pure sciences astronomy, physics, math on top (green) and applied sciences biology and computer science on bottom (blue)

| *Category* | *Closeness* | *Betweenness* | *Normalized* |
|---|---|---|---|
| Astronomy | 2.65 | 575700.5 | 0.09 |
| Astronomy Physics | 2.75 | 407643.3 | 0.04 |
| Quantum Physics | 2.68 | 338569.4 | 0.01 |
| High Energy Physics | 2.68 | 332272.2 | 0.08 |
| Computer Science | 2.91 | 265682.9 | 0.08 |
| Math | 3.18 | 263596.3 | 0.04 |
| General Physics | 2.97 | 219213.4 | 0.05 |

High concentration applied sciences biology and computer science on community with small resolution and pure sciences astronomy, physics and math on Figure 5 can be supported by table of communities with top betweenness values. As we browse through more sparse part of graph on edges we can still see that these nodes interconnect all multi-disciplinary papers with even less degree value.

## 5  Conclusion

Proposed social scholarly network tremendously reinforces the fact that network structure of authors and network structure of audiences are similar.

Multi-disciplinary papers are most mentioned in both in terms of social activity in social network and the most cited in their scholarly network model counterpart.

Social scholarly networks can be a dynamic variety of conventional scholarly networks and by adding new time dimension brings more visibility for researchers in science studies.

**Future Studies**

Integration of scholarly networks with other networks from other domains sound novice idea. For example how can political decision network can interact with scholarly networks. By having multi-relation net work model from diverse domains one can answer or make predictions on how social, political, and cultural values affect scientific research and technological innovation; and even how scientific research and technological innovation affect society, politics, and culture.

Predict emergence of new fields can be fascinating for research institute to make right investments into research topics that a potential for them can be foreseen.

# Refrences